\documentclass[sigplan,screen]{acmart}

\makeatletter
\renewcommand\@formatdoi[1]{\ignorespaces}
\makeatother

\makeatletter
\def\@copyrightspace{\relax}
\makeatother

\NewDocumentCommand{\codeword}{v}{%
\texttt{\textcolor{blue}{#1}}%
}

\AtBeginDocument{%
  }

\setcopyright{none}
\acmISBN{}
\acmYear{}
\acmDOI{}

\acmConference[HotInfra '24]{The 2nd Workshop on Hot Topics in System Infrastructure (HotInfra'24), co-located with SOSP'24}{November 3,
  2024}{Austin, TX, USA}
\begin{document}

\title{Thallus: An RDMA-based Columnar Data Transport Protocol}

\author{Jayjeet Chakraborty}
\email{jayjeetc@ucsc.edu}
\affiliation{%
  \institution{UC Santa Cruz}
  \city{Santa Cruz}
  \state{CA}
  \country{USA}
}

\author{Matthieu Dorier}
\email{mdorier@anl.gov}
\affiliation{%
  \institution{Argonne National Laboratory}
  \city{Lemont}
  \state{IL}
  \country{USA}
}

\author{Philip Carns}
\email{carns@mcs.anl.gov}
\affiliation{%
  \institution{Argonne National Laboratory}
  \city{Lemont}
  \state{IL}
  \country{USA}
}

\author{Robert Ross}
\email{rross@mcs.anl.gov}
\affiliation{%
  \institution{Argonne National Laboratory}
  \city{Lemont}
  \state{IL}
  \country{USA}
}

\author{Carlos Maltzahn}
\email{carlosm@ucsc.edu}
\affiliation{%
  \institution{UC Santa Cruz}
  \city{Santa Cruz}
  \state{CA}
  \country{USA}
}

\author{Heiner Litz}
\email{hlitz@ucsc.edu}
\affiliation{%
  \institution{UC Santa Cruz}
  \city{Santa Cruz}
  \state{CA}
  \country{USA}
}

\renewcommand{\shortauthors}{Chakraborty et al.}

\begin{abstract}
The volume of data generated and stored in contemporary global data centers is experiencing exponential growth. This rapid data growth necessitates efficient processing and analysis to extract valuable business insights. In distributed data processing systems, data undergoes exchanges between the compute servers that contribute significantly to the total data processing duration in adequately large clusters, necessitating efficient data transport protocols. 

Traditionally, data transport frameworks such as JDBC and ODBC have used TCP/IP-over-Ethernet as their underlying network protocol. Such frameworks require serializing the data into a single contiguous buffer before handing it off to the network card, primarily due to the requirement of contiguous data in TCP/IP. In OLAP use cases, this serialization process is costly for columnar data batches as it involves numerous memory copies that hurt data transport duration and overall data processing performance. We study the serialization overhead in the context of a widely-used columnar data format, Apache Arrow, and propose leveraging RDMA to transport Arrow data over Infiniband in a zero-copy manner. We design and implement Thallus, an RDMA-based columnar data transport protocol for Apache Arrow based on the Thallium framework from the Mochi ecosystem, compare it with a purely Thallium RPC-based implementation, and show substantial performance improvements can be achieved by using RDMA for columnar data transport.
\end{abstract}

\maketitle

\section{Introduction}

The past decade has seen a surge in data processing tools and technologies due to the exponential growth of data from IoT devices, sensors, social networks, and financial institutions. Modern data processing systems such as Dask~\cite{rocklin2015dask}, Spark~\cite{Spark}, and BigQuery~\cite{tigani2014google} exchange vast amounts of data over the internal data center network~\cite{stuedi2017crail}, especially during the re-partition and shuffle stages while executing queries. Since these exchanges contribute significantly to the query execution duration, they must exhibit optimal performance.

% talk about the issue in this data transport mechanism
Unfortunately, these data exchanges suffer from serialization overhead, where the in-memory data needs to be serialized into a contiguous representation for the data transport framework to be able to carry it over the wire~\cite{raghavan2021breakfast}~\cite{zerializer}. This is even the case for columnar in-memory formats such as Apache Arrow~\cite{Arrow}, which, although zero-copy within the same host, requires expensive serialization when transferring between hosts. This serialization requirement arises from using TCP/IP-based data transport frameworks since using TCP/IP requires the data to be contiguous before feeding into the network~\cite{camarda1999performance}. In the past decade, I/O became faster, with disk I/O reaching bandwidths of $10$-$15$GB/s with PCIe Gen$5$ NVMe SSDs and network I/O reaching up to $800$ Gbps with the adoption of Terabit Ethernet. With such advancements in I/O performance, data serialization has become a significant source of overhead in network-bound data center applications.

% What can be done to fix it
One solution to this problem is using RDMA~\cite{kalia2016design} to transport columnar data buffers between the machines' memories directly. Although there has been previous work studying the use of RDMA for accelerating network applications and columnar databases~\cite{liu2019design} ~\cite{barthels2015rack}, most of it is either based on research prototypes or used internally in proprietary systems, and none is open-sourced. To our knowledge, no work has been done to explore transporting widely-used open-source in-memory columnar data formats such as Apache Arrow over RDMA.

% what we study and propose
This work studies the serialization overhead incurred when transporting Apache Arrow data over a TCP/IP-based RPC interface implemented using the Thallium~\cite{Thallium} framework from the Mochi~\cite{ross2020mochi} ecosystem. We choose Apache Arrow because it is one of the most widely used columnar in-memory data formats today, and almost all major distributed data processing systems have either adopted it or are working actively to support it. We design and implement Thallus, a data transport protocol based on RDMA to transport Apache Arrow data, leveraging the Thallium framework. We demonstrate that using Thallus to transport columnar data results in up to $5.5$x data transport performance improvement and up to $2.5$x faster query execution over plain Thallium RPCs.

\section{Serialization Overhead in Apache Arrow}
\label{sec:serdes_in_arrow}
To establish our baseline, we built a client-server application that executes SQL queries in the server using DuckDB~\cite{DuckDB} and transports result Arrow record batches over plain Thallium RPCs. Arrow record batches must be serialized into a contiguous buffer to send them through the RPC response. This serialization process incurs copying Arrow column buffers and laying them out contiguously. We execute queries to select all the columns in the dataset and observe that about $30$\% of the RPC duration is spent in serializing a record batch, and only about $0.0004$\% of the duration is consumed in deserialization, as it is a zero-copy operation in Apache Arrow. This work aims to remove the serialization overhead by leveraging RDMA for Arrow data transport. 

\section{Thallus: Design and Implementation}

\begin{figure}[h]
\centering
\includegraphics[width=0.84\linewidth]{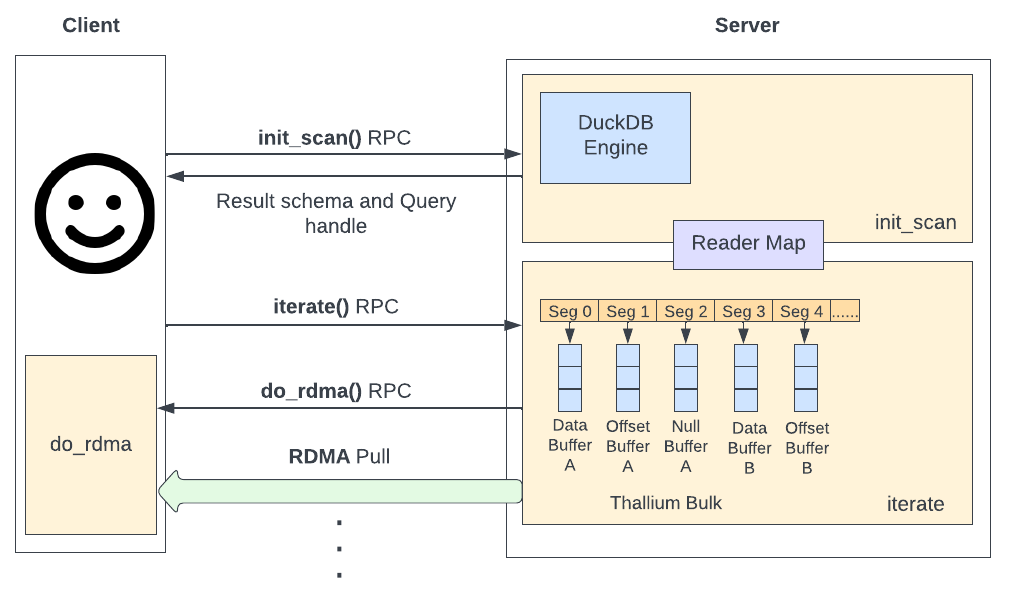}
\caption{Control flow diagram showing the high-level operation of Thallus.}
\label{fig:controlflow}
\end{figure}

Using the Thallium framework, we design and implement Thallus, an RDMA-based data transport protocol for Apache Arrow. Our protocol has $2$ stages: First, we initiate a query execution using a query execution or database engine; Second, we transport the query results batch by batch to the client. We use the Thallium RPC interface to exchange metadata and control information while using the Thallium RDMA interface to transport data buffers. 

Upon connecting to a server, the client invokes an~\codeword{init_scan} procedure on the server with a SQL query and a dataset path as arguments. Inside this procedure, a query execution session is instantiated, and a handle to a result cursor is returned to the client as a response. The client then uses this handle to invoke an~\codeword{iterate} procedure on the server, which iterates over the query cursor, producing results batch by batch, and for every batch, invokes a~\codeword{do_rdma} procedure on the client, that pulls the batches from the server to the client. After all the batches are pulled into the client, a~\codeword{finalize} procedure is invoked to clean up resources and stop the server. Figure~\ref{fig:controlflow} summarizes the high-level operation of Thallus in an implementation with DuckDB as the query execution backend. Next, we discuss each of the procedures mentioned above in detail.

\subsubsection{\textit{Server}: init\_scan}

This procedure takes a SQL query and a dataset path as input arguments. The procedure starts execution by instantiating a~\codeword{DuckDBEngine} instance, a class containing methods for creating and querying a table in DuckDB. We can use a similar interface to leverage any other Arrow-native query execution engine, such as Polars~\cite{Polars} and Velox~\cite{Velox}. Then, a view is created over the dataset, the SQL query is parsed, and an optimized physical plan is generated for execution. Upon initiating the query execution, DuckDB returns a query cursor instance. Arrow provides a~\codeword{RecordBatchReader} interface for streaming access to Arrow record batches. We extend this interface to create a~\codeword{DuckDBRecordBatchReader} interface that, upon every iteration, uses the DuckDB query cursor to generate DuckDB data chunks and converts them to equivalent Arrow record batches. We achieve this conversion in a zero-copy manner using the Arrow C Data Interface~\cite{cdata}. Finally, we store a pointer to this reader instance in a map, which we refer to as the reader map, using a unique UUID, read the result's schema from the reader instance, and send both back to the client as a response.

\subsubsection{\textit{Server}: iterate}

This procedure is used to iterate over the~\codeword{DuckDBRecordBatchReader} and produce result Arrow record batches. It takes the record batch reader UUID as its only input argument to identify a particular query execution instance in a multi-tenant environment. It starts execution by looking up the UUID in the reader map to find the reader instance for the client. It then iterates over the reader to produce result Arrow record batches, that are sent to the client over RDMA.

For the RDMA operations, we use the Thallium "bulk" interface, as it provides access to remote memory regions. The users must allocate a list of segments, where each segment points to a distinct memory region. These segments are then prepared for remote reads and write through an "expose" operation, which generates a bulk handle. The bulk handle is a descriptor for an RDMA-ready pinned remote memory region that can be serialized and passed around through RPC arguments and responses and read for accessing remote memory regions.

In Apache Arrow, a record batch is a collection of columns, where each column is composed of $3$ buffers representing the data values, offsets, and null masks. We allocate $3$x the number of segments as there are columns and for every $i$ th column, we map its data, offset, and null buffer to the $3i$ th, $3i+1$ th, and $3i+2$ th segment, respectively. While mapping the buffers to the segments, we keep storing the sizes of each of the data, offset, and null buffers in $3$ separate vectors. We expose these segments as a read-only Thallium bulk and invoke the client-side~\codeword{do_rdma} procedure with the bulk handle, the buffer sizes vectors, and the number of rows in the batch. The buffer size vectors must be shipped to the client for it to allocate a similar layout of buffers as on the server to pull the server-side buffers into the client-side ones in a one-to-one manner. When the reader is exhausted and no more batches can be produced, the procedure returns.

\subsubsection{\textit{Server}: finalize}
The~\codeword{finalize} procedure is invoked after the client has finished pulling all the batches from the server and needs to terminate the data transport operation. This procedure frees the allocated buffers and the reader map and finalizes the Thallium engine.

\subsubsection{\textit{Client}: do\_rdma}
This procedure pulls the Arrow record batches from the server to the client. It is invoked by the server-side~\codeword{iterate} procedure with the number of rows, vectors of buffer sizes, and the remote bulk handle as arguments. It then uses this information to allocate buffers for the client to store data fetched from the server and exposes these buffers as a write-only local bulk. It then pulls the remote bulk into the local bulk using RDMA. The bulk pulling is a scatter-gather operation where the server-side buffers are fetched into the corresponding client-side buffers. Once all the buffers are brought into the client, they are associated with buffer sizes and data types (already available on the client) to instantiate column structures. These column structures are then wrapped with the result schema to create an Arrow record batch. The record batches are then written to an output stream, such as a parquet file.

\section{Evaluation}
We benchmark Thallus by executing SQL queries that generate result sets of different sizes and use a pure Thallium RPC-based implementation as our baseline. We measure the transport and end-to-end query execution duration using Thallus and compare it with our baseline. We measured the transport duration only by eagerly collecting all the results in the server memory first and then letting the client read them.

\begin{figure}[h]
\centering
\includegraphics[width=0.80\linewidth]{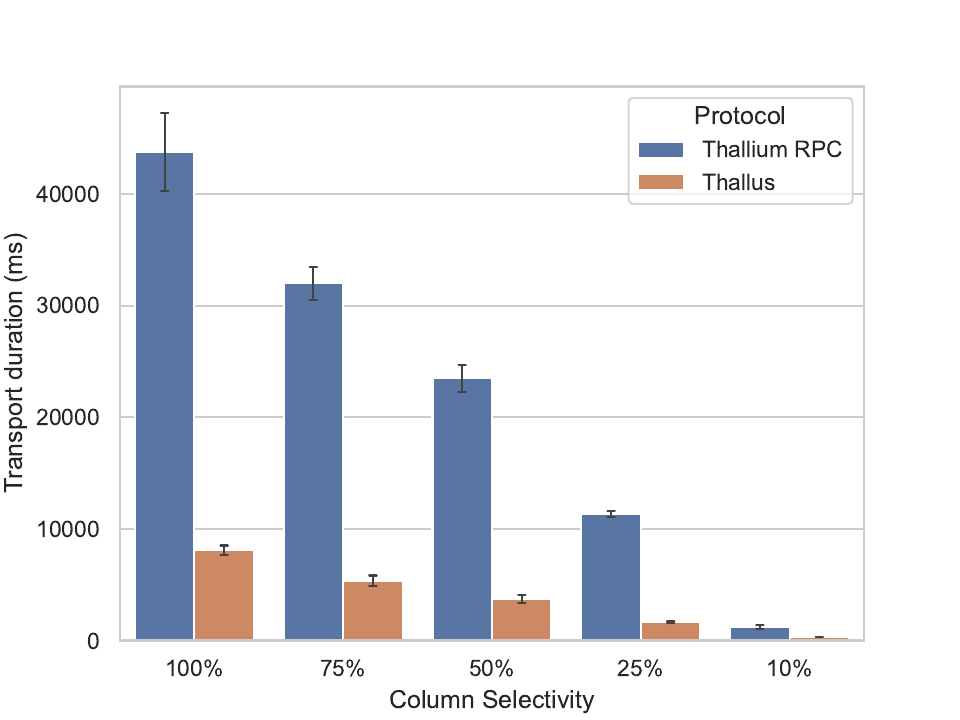}
\caption{Comparison of data transport duration using Thallus versus Thallium RPC in column selectivity experiments.}
\label{fig:colq}
\end{figure}

\begin{figure}[h]
\centering
\includegraphics[width=0.80\linewidth]{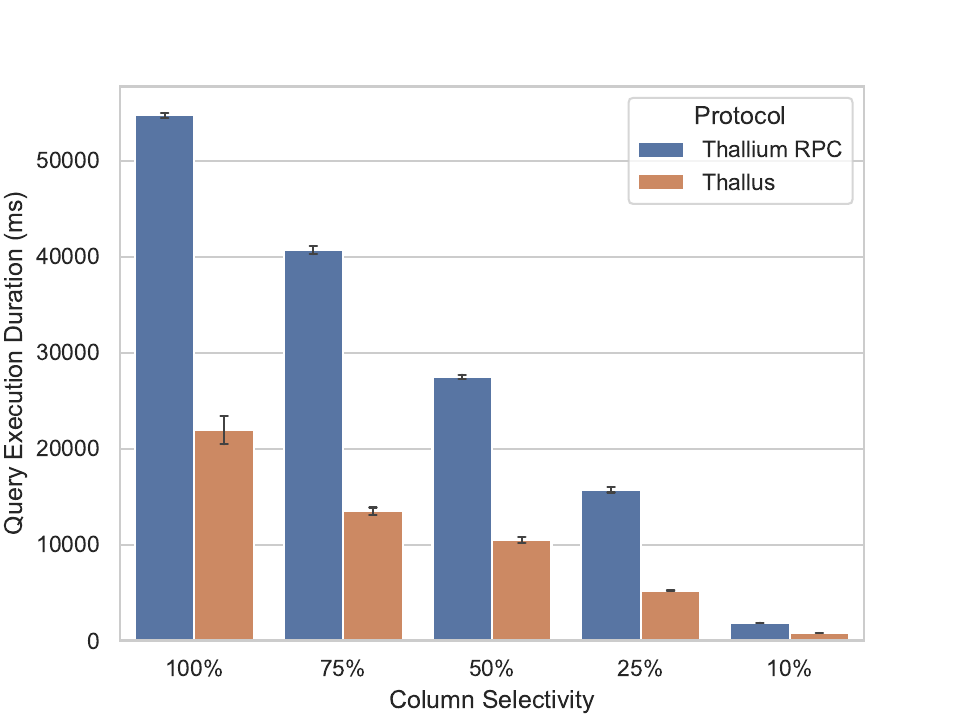}
\caption{Comparison of query execution duration using Thallus versus Thallium RPC in column selectivity experiments.}
\label{fig:colqtotal}
\end{figure}

As shown in Figure~\ref{fig:colq} and Figure~\ref{fig:colqtotal}, we observe that Thallus is up to $5.5$x faster than Thallium RPC in data transport durations and achieves up to $2.5$x performance gain in end-to-end query execution duration. As can be seen from both figures, the relative performance gain of Thallus over Thallium RPC diminishes with the reduction in the size of the result set. This is because when the batch size transferred reduces, the constant overheads of doing RDMA, such as buffer allocation and memory pinning, become significant. 

\section{Conclusion}
In this paper, we explore using RDMA to accelerate the transport of columnar data in distributed data processing systems. We begin by measuring the overhead caused by the expensive serialization of Apache Arrow data when transported over the wire using a traditional TCP/IP-based RPC framework. To tackle the serialization overhead, we design and implement Thallus, an RDMA-based columnar data transport protocol using the Thallium framework from the Mochi ecosystem to transport Apache Arrow data over Infiniband. Internally, Thallus uses RPCs for control operations and RDMA for data operations. We structure our implementation as a client-server model where queries are executed in a DuckDB instance running on the server, and results are sent back to the client using Thallus. We benchmark Thallus against a pure RPC-based implementation, which was also written using the Thallium framework. Thallus exhibits up to $5.5$x improvement in data transport performance and up to $2.5$x improvement in end-to-end query execution performance over TCP/IP-based Thallium RPCs on OLAP-style columnar workloads. Our studies show that RDMA can be a viable and more performant alternative to TCP/IP for data transport in modern data processing applications.

\bibliographystyle{ACM-Reference-Format}
\bibliography{sample-base}

\end{document}